\documentclass[fleqn,usenatbib]{mnras}

\usepackage{newtxtext,newtxmath}

\usepackage[T1]{fontenc}
\usepackage{ae,aecompl}

\usepackage{graphicx}	
\usepackage{amsmath}	
\usepackage{amssymb}	

\title[Dark matter mass function from lensing]{Constraining sterile neutrino cosmologies with strong gravitational lensing observations at redshift $z\sim$ 0.2}

\author[S. Vegetti et al.]{
S. Vegetti,$^{1}$\thanks{E-mail: svegetti@mpa-garching.mpg.de}
G. Despali,$^{1}$
M. R. Lovell,$^{2,3,4}$
and
W. Enzi$^{1}$
\\
$^{1}$ Max-Planck Institute for Astrophysics, Karl-Schwarzschild Str. 1, D-85748, Garching, Germany\\
$^{2}$ Center for Astrophysics and Cosmology, Science Institute, University of Iceland, Dunhagi 5, 107 Reykjavik, Iceland \\
$^{3}$ Institute for Computational Cosmology, Durham University, South Road, Durham DH1 3LE, UK\\
$^{4}$ Max-Planck-Institut f\"ur Astronomie, K\"onigstuhl 17, D-69117 Heidelberg, Germany\\
}

\date{Accepted XXX. Received YYY; in original form ZZZ}

\pubyear{2015}

\begin{document}
\label{firstpage}
\pagerange{\pageref{firstpage}--\pageref{lastpage}}
\maketitle

\begin{abstract}
We use the observed amount of subhaloes and line-of-sight dark matter haloes in a sample of 11 gravitational lens systems from the Sloan Lens ACS Survey to constrain the free-streaming properties of the dark matter particles. In particular, we combine the detection of a small-mass dark matter halo by \citet{Vegetti10} with the non-detections by \citet{Vegetti14} and compare the derived subhalo and halo mass functions with expectations from cold dark matter (CDM) and resonantly produced sterile neutrino models. 
We constrain the half-mode mass, i.e. the mass scale at which the linear matter power spectrum is reduced by 50 per cent relatively to the CDM model, to be $\log M_{\rm{hm}} \left[M_\odot\right]  < 12.0$ (equivalent thermal relic mass $m_{\rm th} > 0.3$ keV) at the 2$\sigma$ level. This excludes sterile neutrino models with neutrino masses $m_{\rm s} < 0.8$ keV at any value of $L_{\rm 6}$. Our constraints are weaker than currently provided by the number of Milky Way satellites, observations of the 3.5 keV X-ray line, and the Lyman $\alpha$ forest. However, they are more robust than the former as they are less affected by baryonic processes. Moreover, unlike the latter, they are not affected by assumptions on the thermal histories for the intergalactic medium. 
Gravitational lens systems with higher data quality and higher source and lens redshift are required to obtain tighter constraints.
 \end{abstract}




\section{Introduction}

A significant number of astrophysical observations supports the widespread presence of dark matter and its dominant contribution to the matter content of the Universe.
However, the nature of dark matter is to this day still an unsolved problem. In the standard $\Lambda$CDM cosmological model, dark matter is assumed to be cold, i.e. to have had negligible thermal velocity at early times, and possible candidates include the neutralino and the axions \citep[e.g.][]{Bottino03,Ringwald16}. While this model is very successful at reproducing large-scale observations, some discrepancies arise at galactic and sub-galactic scales \citep[e.g.][]{Moore1994, Klypin1999, Kuzio2008, deBlok2010, Walker2011, Amorisco2012, Boylan2012}.

Concerning these issues, alternative models have been considered \citep[e.g.][]{Lovell14,Vog16,Robles17,Irsic17b}, among which are the so-called warm dark matter models, where the corresponding particles have non-negligible thermal velocities. In particular, sterile neutrinos with masses of a few keV have been shown to be an interesting candidate for warm dark matter particles \citep[e.g.][and references therein]{Boyarsky12}. Moreover, the three sterile neutrino model, also known as the neutrino minimal standard model ($\nu$MSM), can also explain baryogenesis processes and neutrino oscillations \citep[][and references therein]{Boyarsky12}. 
Recently, sterile neutrinos have gained renewed attention following the apparent detections of a 3.5 keV X-ray line in several galaxy clusters as well as M31, and at the centre of the Milky Way \citep{Boyarsky14, Bulbul14, Boyarsky15}; for a review see, e.g. \citet{Iakubovskyi14}. However, sterile neutrino decay as a source of the line remains contentious \citep[e.g.][]{Anderson14,Gu15,Jeltema16}. Recent results from the Hitomi collaboration \citep{Aharonian16} have ruled out an atomic transition as the origin of the 3.5 keV line, but a dark matter interpretation is still possible due to the different sensitivity of the Hitomi instrument to narrow (atomic) and broad lines. The 11$\sigma$ detection of the 3.5 keV signal by NuStar \citep{Neronov16} from a quiescent region of the sky and subsequent confirmation of this result with Chandra \citep{Cappelluti17} suggest that the line is not a statistical fluctuation nor an artefact of the XMM-Newton's instrumentation. It is therefore important to find complimentary probes of these models as well as supporting future X-ray observatory missions such as XARM and ATHENA that will provide still stronger X-ray constraints.

Due to their appreciable thermal velocities at early times, warm dark matter particles free-stream out of density perturbations and are responsible therefore for a cut-off in the linear power spectrum and by extension in the formation of structures at small scales. The exact scale at which the suppression happens is strongly related to the dark matter particle production mechanism. For example, for 7 keV sterile neutrinos with a lepton asymmetry (defined in Section \ref{sec:dm_model}) much smaller or larger than $L_6 = 8$  this happens at the scale of progressively more massive dwarf galaxies \citep{Lovell16}, while in a CDM model the suppression happens at much smaller scales. Quantifying the number of small mass structures is therefore an essential approach to distinguish among different dark matter models. For example, \citet{Kennedy14} and \citet{Lovell16} have used semi-analytic models to compare the abundance of luminous Milky Way satellites predicted by sterile neutrino dark matter models with the observed number of dwarf spheroidals, and derived constraints on the sterile neutrino mass and the lepton asymmetry $L_6$. However, since luminous satellites are a biased tracer of the underlying mass distribution, the interpretation of these results is complicated by assumptions made on baryonic processes. Moreover, the not so precisely known mass of the Milky Way is also a source of degeneracy in the number of expected satellites. 

Recently, measurements of the Lyman $\alpha$ forest from high redshift QSO spectra have argued that all values of  $L_6$ for sterile neutrino mass $<$10~keV are ruled out. However, these results are highly sensitive to assumptions made on the thermal histories for the inter-galactic medium \citep{Baur17}. It has been demonstrated by \citet{Garzilli17} that if one does not make any assumption about the (experimentally unknown) state of the intergalactic medium at $z > 5$, these bounds get considerably relaxed. As a result, the Lyman $\alpha$ limits are less stringent and in particular, the 7 keV sterile neutrino model, consistent with the 3.5 keV line detection, are within the allowed range \citep{Baur17}.

Strong gravitational lensing, being sensitive to gravity, provides a more direct method to quantify the dark matter distribution at sub-galactic scales and in a way which is less affected by baryonic processes \citep[however, see][]{Despali17}. \citet{Mao98} first proposed the idea to constrain the amount of small-mass structure in gravitational lens galaxies via their influence on the relative fluxes of multiply-imaged quasars. 
Subsequently, \citet{Dalal02} were the first to use a relatively small sample of seven gravitationally lensed quasars to derive a statistical constraint on the projected mass fraction in substructure that is larger than, but consistent with, predictions from CDM numerical simulations. At the same time, analyses of CDM numerical simulations \citep{Xu09,Xu15} have shown that the predicted substructure population is not sufficient to reproduce the level of flux-ratio anomalies currently observed, and that complex baryonic structures in the lens galaxies are a likely explanation \citep[see also][]{McKean07,Gilman17,Hsueh16,Hsueh17b,Hsueh17a}. In the future, narrow-line observations for a large number of quadruply imaged quasars, which are expected to be discovered by large-scale surveys, will allow one to investigate these issues further and potentially provide new interesting constraints on the properties of dark matter \citep{Nierenberg17,Gilman18}.

In the meantime, \citet{Koopmans05} and \citet{Vegetti09a} have introduced the \emph{gravitational imaging} technique that uses magnified arcs and Einstein rings to detect and measure the mass of individual subhaloes \citep[see also][]{Vegetti10a}. 
Unlike other methods, this technique treats substructures not as analytical mass clumps but as pixellated linear potential corrections to the main lensing potential. As such, it does not require any prior assumption on the number of substructures nor their density profile and redshift. Moreover, it can easily distinguish a substructure from a smooth, but complex, mass distribution \citep{Barnabe09,Vegetti14} and is therefore less prone to false detections. However, unlike studies of flux-ratio anomalies, it relies on increasingly high-resolution data for the detection of the smaller mass haloes. Using the gravitational imaging technique \citet{Vegetti10} and \citet{Vegetti12} have reported the detection of two small mass substructures, while \citet{Vegetti14} have used a sample of 11 lenses to derive statistical constraints on the CDM substructure mass function, which are consistent with CDM predictions.
 
At the other end of the spectrum, several methods have been developed to detect mass substructures not individually, but via their collective gravitational effect \citep{Fadely12, Hezaveh16, Cyr-Racine16, Birrer17, Daylan17, Chatterjee17}. Understanding the biases introduced by complex mass distributions of the main lensing galaxy, as highlighted by \citet{Hsueh16,Hsueh17b,Hsueh17a}, \citet{Gilman17}, and \citet{Vegetti14}, will be critical for a practical application of these techniques. 

Most if not all analyses so far have considered subhaloes within lens galaxies as the only source of perturbation to the lensed images. However, as demonstrated by \citet{metcalf05} and \citet{Despali18}, the contribution from line-of-sight haloes can be significant and even dominant, depending on the lens and source redshifts. Moreover, both the number and the structure of line-of-sight haloes are less affected by feedback and accretion processes and can therefore be used to get tighter and cleaner constraints on the properties of dark matter. 

In this paper we reanalyse the sample by \citet{Vegetti14} to include the contribution of both subhaloes and line-of-sight haloes to the total number of detectable objects. 
In particular, we combine the non-detections by \citet{Vegetti14} and the detection by \citet{Vegetti10}  to derive statistical constraints on the halo and subhalo mass functions. As the key parameters that are set by the dark matter model are common to both mass functions, the inclusion of line-of-sight haloes represents an increase in the constraining power. This paper is organized as follows. The data is introduced in Section \ref{sec:data}. In Section \ref{sec:dm_model} we present the sterile neutrino dark matter model under consideration. In Sections \ref{sec:mass} and \ref{sec:mass_function} we provide all the definitions of mass for subhaloes and line-of-sight haloes as well as the mass function expressions adopted for both populations. In Sections \ref{sec:likelihood} and \ref{sec:posterior} we introduce the complete likelihood function, the model parameters, and the corresponding priors. Finally, our results are presented and discussed in Sections \ref{sec:results} and \ref{sec:conc}, respectively.

\section{Data}
\label{sec:data}

The analysis presented in this paper is based on a sample of 11 gravitational lens systems taken from the SLACS survey \citep[][and references therein]{Bolton06, Auger10}. A summary of the systems with the lens and source redshift can be found in Table \ref{tbl:lenses}. 
\citet{Vegetti10} and \citet{Vegetti14} have previously modelled this sample with the Bayesian grid-based adaptive code by \citet{Vegetti09a}.
 To summarize, \citet{Vegetti10} have reported the detection of a dark-matter-dominated perturber that was then interpreted as a substructure with a total mass of $3.5\times10^9M_\odot$ (under the assumption of a Pseudo-Jaffe profile located on the plane of the host lens, hereafter PJ, and corresponding roughly to $M_{\rm vir}^{\rm NFW}\sim10^{10}M_\odot$) in the lens system SDSSJ0946+1006. Later on, \citet{Vegetti14} have shown that no substructure is required for all the other lenses in the sample. For each lens, they have also derived the so-called sensitivity function, that is the smallest detectable substructure mass at each relevant position on the image plane.  Here, we make use of their results as an input to our analysis and refer to their papers for more details on the data and how these results were obtained. In particular, we relax the assumption that all perturbers are substructures and allow for the possibility of a contribution from line-of-sight haloes, for the statistical interpretation of both detections and non-detections.
 
\begin{table}
\begin{center}
\caption{The list of the gravitational lens systems considered in this paper, with the lens and source redshifts.}
\begin{tabular}{lll}
\hline
Name (SDSS)&$z_{\rm{lens}}$&$z_{\rm{source}}$\\
\hline

$\rm{J0252+0039}$&0.280&0.982\\

$\rm{J0737+3216}$&0.322&0.581\\

$\rm{J0946+1006}$&0.222&0.609\\

$\rm{J0956+5100}$&0.240&0.470\\

$\rm{J0959+4416}$&0.237&0.531\\

$\rm{J1023+4230}$&0.191&0.696\\

$\rm{J1205+4910}$&0.215&0.481\\

$\rm{J1430+4105}$&0.285&0.575\\

$\rm{J1627-0053}$&0.208&0.524\\

$\rm{J2238-0754}$&0.137&0.713\\

$\rm{J2300+0022}$&0.228&0.463\\
\hline
\end{tabular}
\label{tbl:lenses}
\end{center}
\end{table}


\section{Dark matter model}
\label{sec:dm_model}

The dark matter particle considered in this paper is the resonantly produced sterile neutrino \citep{Shi99,Dolgov:00,Asaka05,Laine08,Boyarsky09a}. In this section, we present a summary of the model and its application in astronomy. For an in-depth description of both, we refer the interested reader to \citet{Lovell16}.

The sterile neutrino is produced at high energies, around the gluon-hadron transition ($\sim$10~MeV). It originates from the oscillation of active neutrinos, and the probability of this oscillation is enhanced in the presence of a lepton asymmetry, i.e. an excess of leptons over anti-leptons. It is parametrized using the $L_6$ parameter: $L_{6}=10^{6}(n_\rmn{l}-\bar{n}_\rmn{l})/s$, where $n_\rmn{l}$ is the number density of leptons, $\bar{n}_\rmn{l}$ the number density of anti-leptons, and $s$ the entropy density. The ability to enhance the production of sterile neutrinos via lepton asymmetry reduces the required mixing angle between sterile neutrinos and active neutrinos to obtain the measured universal dark matter density, and thus evades bounds on sterile neutrino parameters derived from X-ray observations.

A second important consequence of lepton asymmetry-induced production is the effect on the sterile neutrino velocities. This is because the enhancement in the production rate due to the lepton asymmetry scales with the momentum of the neutrinos. 

Small asymmetries  ($L_6<6-25$, depending on the sterile neutrino mass) lead to excess production of low momentum sterile neutrinos and are therefore `cooler' distributions than the no asymmetry case. At higher asymmetries, the enhancement in production is stronger and extends to high momentum sterile neutrinos. For the maximal lepton asymmetry ($L_{\rm 6} > 100$) all momenta receive an equal boost compared to the zero asymmetry scenario and the resulting momentum distribution is therefore almost identical to that of no lepton asymmetry. Given that the mass of the sterile neutrino also plays a role in the momentum distribution, with less massive particles exhibiting higher velocities, the free-streaming scale, and thus the halo mass function, is specified by two parameters: sterile neutrino mass $m_\rmn{s}$, and lepton asymmetry $L_6$.

The momentum distribution functions are calculated using the methods of \citet{Laine08} and are used as inputs for Boltzmann solver codes to calculate the linear matter power spectrum, $P(k)$. In our case, we use matter power spectra that have been computed using a modified version of {\sc camb} \citep{Lewis00}, and examples of these are displayed in Fig.~4 of \citet{Lovell16}. These curves exhibit an array of cutoff slopes and positions. For this study we characterise each curve by the transfer function of the sterile neutrino linear matter power-spectrum compared to its CDM counterpart, which is given by $T(k)=[P(k)_\rmn{SN}/P(k)_\rmn{CDM}]^{0.5}$, and specifically the wavenumber at which the transfer function has the value 0.5. This is known as the half-mode wavenumber, $k_\rmn{hm}$. Its influence on the halo mass function is parametrised through a dependent parameter, the half-mode mass $M_\rmn{hm}$, which takes the following form:
\begin{equation}
    M_\rmn{hm}=\frac{4\pi}{3}\bar{\rho}\left(\frac{\pi}{k_\rmn{hm}}\right)^{3}\,,    
\end{equation}
where $\bar{\rho}$ is the present day mean matter density of the Universe. In practice, $M_\rmn{hm}$ is a function of both $m_\rmn{s}$ and $L_{6}$, thus placing constraints on the former will lead to limits on the sterile neutrino parameters. In some cases $M_\rmn{hm}$ does not fully encapsulate the fine details of the matter power spectra, such as the shallower slopes of some models: we comment below on how a more accurate parametrisation of these curves would likely change the results.


\section{Mass definition}
\label{sec:mass}

We assume the \emph{true} mass of line-of-sight haloes and substructures, $m$, to be the virial mass of a \citet[][hereafter NFW]{Navarro97} profile with a \citet{Duffy08} concentration-mass relation. While this is a good description for the former, it is only an approximation for the latter. \citet{Despali18} have shown that at fixed virial mass these have a larger concentration that is mildly dependent on the subhalo distance from the host centre. However, by comparing the different deflection angles, they have also shown that assuming a constant  \citet{Duffy08} mass-concentration relation plays a secondary role in terms of the lensing effect. In particular, this assumption leads to an error on the mass that is as low as 5 per cent for subhaloes with masses of 10$^{5-6}M_\odot$ and as large as 20 per cent for masses of $10^{9}M_\odot$; the error on the expected number of substructure is of the order of 10 per cent. We also assume that the concentration does not change with the dark matter model. As shown by \citet{Ludlow16} the concentration of WDM haloes differs from the CDM  case only at low masses, where the number of structures is strongly suppressed. Again, \citet{Despali18} have shown that this assumption is of secondary importance. We refer to their paper for a more detailed discussion on the matter. 

We assume that the \emph{observed} mass $m^{\rm o}$ of perturbers, i.e. the mass that one would derive from the gravitational lens modelling of the data, and the lowest detectable mass are PJ total masses located on the plane of the host lens. We then use the mass-redshift relation derived by \citet{Despali18} to statistically relate the \emph{true} and \emph{observed} masses to each other as described in Section \ref{sec:exp_val}.
This approach follows from the fact that while the detections were made in a pixellated model independent way, they have been then characterised in terms of PJ substructures.
The sensitivity function used in this paper and derived by \citet{Vegetti14} has also been obtained under the assumption of PJ substructures. 


\section{Dark matter mass function}
\label{sec:mass_function}

While \citet{Vegetti14} have assumed a CDM model and have focused only on substructures, here, we allow for a more general dark matter model that includes the effect of particle free-streaming and the contribution from small-mass dark matter haloes located along the line of sight. Following \citet{Lovell14} and \citet{Schneider12}, we parametrize the substructure and the halo mass function as follows:
\begin{equation}
n(m) = n(m)^{\rm CDM}\left(1+\frac{M_{\rm hm}}{m}\right)^{\beta}\,,
\label{eq:mass_f}
\end{equation}
where the second factor expresses the effect of particle free-streaming. A slightly different parametrization was found to be a better fit for the subhalo mass function by \citet{Lovell14}:
\begin{equation}
n(m) = n(m)^{\rm CDM}\left(1+\frac{2.7\times M_{\rm hm}}{m}\right)^{\beta}\,.
\label{eq:mass_f2}
\end{equation}
In this paper, we make use of both, with the first one leading to a more significant number of WDM subhaloes and hence being more conservative. We assume the CDM mass function for substructures to be
\begin{equation}
n_{\rm sub}^{\rm CDM}(m) \propto  m^{-\alpha}\,.
\label{eq:sub_mf}
\end{equation}
For the line-of-sight halo CDM mass function we adopt the expression by \citet{Sheth99b}, with the best-fitting parameters optimized for the Planck cosmology
calculated by \citet{Despali16}. As discussed in the previous section, for both populations the mass function is a function of the NFW virial mass for a fixed concentration-mass relation. Scatter in this relation is taken into account in a statistical sense as described in Section \ref{sec:exp_val}.


\section{Likelihood function}
\label{sec:likelihood}

In this section, we derive an expression for the likelihood of detecting $n$ perturbers (substructures plus line-of-sight haloes) with observed masses $\left\{m^{\rm ob}_1, ....,m^{\rm ob}_n\right\}$ at the projected positions  $\left\{{\bf x}^{\rm ob}_1, ....,{\bf x}^{\rm ob}_n\right\}$, and no detection in all other mass and position ranges.  Assuming a Poisson distribution we follow \citet{Marshall83} and write the log-likelihood for a single  lens galaxy as follows (see Appendix \ref{sec:likel} for a derivation):
\begin{multline}
\log P\left(\left\{m^{\rm ob}_1, ....,m^{\rm ob}_n\right\},\left\{{\bf x}^{\rm ob}_1, ....,{\bf x}^{\rm ob}_n\right\}|\theta\right) \\ = 
-\int{\left[\mu_s(m^{\rm o},\mathbf{x}^{\rm o})+\mu_l(m^{\rm o},\mathbf{x}^{\rm o})\right]dm^{\rm o}d\mathbf{x}^{\rm o}}\\+
\sum_i^n \log\left[\mu_s\left(m_i^{\rm ob},\mathbf{x}^{\rm ob}_i\right)dm^{\rm o}d\mathbf{x}^{\rm o}+\mu_l\left(m_i^{\rm ob},\mathbf{x}^{\rm ob}_i\right)dm^{\rm o}d\mathbf{x}^{\rm o}\right]\,.
\label{eq:likelihood}
\end{multline}
Here, $\theta$ is a vector containing the model free parameters that define the (sub)structure mass function, and is explicitly introduced in Section \ref{sec:posterior}. $\mu_s \left(m^{\rm o},\mathbf{x}^{\rm o}\right)dm^{\rm o}d\mathbf{x}^{\rm o}$  and  $\mu_l\left(m^{\rm o},\mathbf{x}^{\rm o}\right)dm^{\rm o}d\mathbf{x}^{\rm o}$ are the expected number of substructures and line-of-sight haloes, respectively, in the mass range $m^{\rm o},m^{\rm o}+dm^{\rm o}$ and projected position range ${\bf x}^{\rm o},{\bf x}^{\rm o}+ d{\bf x}^{\rm o}$. A derivation of them is given in Section \ref{sec:exp_val}. The masses $m^{\rm o}$ in the above equations are intended as \emph{observed} ones and are defined according to Section \ref{sec:mass}. These are integrated between the lowest detectable mass at each projected position, $M^{\rm PJ}_{\rm low}(\mathbf{x}^{\rm o})$, and $M^{\rm PJ}_{\rm max}=1.0\times10^{10}~M_\odot$. Both limits are intended as PJ total masses. It should be noted that here $\mathbf{x}^{\rm o}$ is the position on the plane of the main lens where the perturber can be detected, for substructures this corresponds with the projected position of the perturber, within a relatively small error. For line-of-sight haloes, due the multiple lens plane configuration, $\mathbf{x}^{\rm o}$ and $\mathbf{x}$ are related to each other via the recursive lens equation evaluated at $\mathbf{x}^{\rm o}$.

\subsection{Sensitivity function}
\label{sec:sens}

For each of the considered lenses, the substructure sensitivity function, that is the lowest detectable mass as a function of position on the image plane $M^{\rm PJ}_{\rm low}(\mathbf{x}^{\rm o})$, was derived by \citet{Vegetti14}. Briefly, this was calculated by identifying the smallest PJ total mass on the plane of the host lens responsible for a change of the Bayesian evidence by  $\Delta\log E\leq-50$, relatively to a model with no substructure. Under the assumption of Gaussian noise, this corresponds to a 10$\sigma$ detection. As a reference, the detections reported by \citet{Vegetti10} and \citet{Vegetti12} were, respectively, at the 16$\sigma$ and 12$\sigma$ limit.
As demonstrated by \citet{Vegetti14} and \citet{Despali18}, considering a constant sensitivity across the image plane can have a significant impact on the expected number of detectable objects and therefore on the inferred mass function parameters. Thus, we have derived the sensitivity function for each pixel on the image plane with a signal-to-noise ratio (SNR) larger than three. In the following sections, we include its effect via $P(I=1|m^{\rm o},\mathbf{x}^{\rm o})$, which is expressed as
\begin{equation}
P(I=1|m^{\rm o},\mathbf{x}^{\rm o}) = \begin{cases} 1 & \mbox{if} ~m^{\rm o} \geq M_{\rm low}^{\rm PJ}(\mathbf{x}^{\rm o}) \\ 0 & \mbox{otherwise} \end{cases}\,,
\end{equation}
where $I$ is a vector which is equal to one for detectable perturbers and zero otherwise. By definition, $P(I=1|m^{\rm o}=m_i^{\rm ob},\mathbf{x}^{\rm o} =\mathbf{x}_i^{\rm ob}) =1$.

\subsection{Expectation values}
\label{sec:exp_val}

We now provide explicit expressions for the expectation values of substructures and line-of-sight haloes. Below, the integration limits are intended between ${M^{\rm NFW}_{\rm min}}=10^5M_\odot$ and ${M^{\rm NFW}_{\rm max}}=10^{11}M_\odot$ for the \emph{true} NFW virial mass and within the SNR $\geq 3$ region for the true projected position. For the redshift of line-of-sight haloes we integrate between the observer and the source, but exclude the region within the virial radius of the main lens, i.e. $z\in\left[z_{\rm lens}-10^{-4};z_{\rm lens}+10^{-4}\right]$. 
The substructure expectation value is given by
\begin{multline}
\mu_s(m^{\rm o},\mathbf{x}^{\rm o}) = \mu_{0,s} \times\int P(I=1|m^{\rm o},\mathbf{x}^{\rm o})P(m^{\rm o}|m,z_{\rm lens})\\P(m|\theta)P(\mathbf{x}^{\rm o}|\mathbf{x},z_{\rm lens})P(\mathbf{x})~dmd\mathbf{x}\,,
\end{multline}
where 
\begin{equation}
P(m|\theta) = n(m|\theta)\left[\int{ n(m^\prime|\theta)dm^\prime}\right]^{-1}
\end{equation}
is such that $P(m|\theta)dm$ is the probability of finding one substructure in the mass range $m,m+dm$.

Introducing the projected dark matter mass of the primary lens $M_{\rm lens}$ within the region of interest and the corresponding projected dark matter mass fraction in substructure $f_{\rm sub}$ we can express $\mu_{0,s}$ as follows
\begin{equation}
\mu_{0,s} = f_{\rm sub} M_{\rm lens}\left[\int {m^\prime P(m^\prime|\theta)dm^\prime}\right]^{-1}\,.
\label{equ:mu_s}
\end{equation}
In this paper, we define $f_{\rm sub}$ as the projected dark matter fraction in substructure with masses between ${M^{\rm NFW}_{\rm min}}$ and ${M^{\rm NFW}_{\rm max}}$ and within the considered region. 
In particular, $f_{\rm sub}$ is a mean value and is, therefore, the same for every galaxy in the sample. This assumption is not critical here, as we are considering a sample of lenses which is relatively homogeneous both in mass and redshift. This definition differs from the one by \citet{Vegetti14}, as it uses a different definition of substructure mass, as well as a different parametrization of the substructure mass function.

Similarly, the expectation value for line-of-sight haloes is given by
\begin{multline}
\mu_l(m^{\rm o},\mathbf{x}^{\rm o}) = \mu_{0,l} \times\int P(I=1|m^{\rm o},\mathbf{x}^{\rm o})P(m^{\rm o}|m,z)\\P(m,z |\theta)P(\mathbf{x}^{\rm o}|\mathbf{x},z)P(\mathbf{x})P(z)~dmdz d\mathbf{x} \,,
\label{equ:mu_l}
\end{multline}
where $\mu_{0,l}$ is expressed as 
\begin{equation}
\mu_{0,l} = \int n(m^\prime,z^\prime|\theta)\frac{dV(\mathbf{x}^\prime)}{dz^\prime}dm^\prime dz^\prime d\mathbf{x}^\prime\,.
\end{equation}
$P(m,z |\theta)dmdz$ is the probability of finding one line-of-sight halo in the mass range $m,m+dm$ and in the redshift range $z,z+dz$, and is related to the halo mass function as follows:
\begin{equation}
P(m,z|\theta)dmdz = n(m,z|\theta)\frac{dV}{dz}dmdz\left[\int n(m^\prime,z^\prime|\theta)\frac{dV}{dz^\prime}dm^\prime dz^\prime\right]^{-1}\,.
\end{equation}
As shown by \citet{Despali18}, the measurement error on the perturber positions is relatively small (i.e. within 2.5 times the PSF FWHM), hence for simplicity we assume $P(\mathbf{x}^{\rm o}|\mathbf{x},z)=\delta(\mathbf{x}-g(\mathbf{x}^{\rm o},z))$. For substructure $g(\mathbf{x}^{\rm o},z)\equiv \mathbf{x}^{\rm o}$, for line-of-sight haloes $g(\mathbf{x}^{\rm o},z)$ takes into account the effect of the recursive lens equation. Following the results by \citet{Xu15} and \citet{Despali17} we assume a uniform probability for $P(\mathbf{x})$. 

By comparing the lensing effect of PJ perturbers at the redshift of the main lens with those of NFW line-of-sight haloes and subhaloes, \citet{Despali18} have derived a mass-redshift relation that allows one to map one population into the other; following their results and referring to the mass-redshift relation as $f(m,z)$ we define $P(m^{\rm o}|m,z)$ as follows:
\begin{equation}
P(m^{\rm o}|m,z) = \frac{1}{\sqrt{2 \pi}m^{\rm o}\sigma(z)}\exp{\left[-\frac{\left(\log m^{\rm o}-f(m,z)\right)^2}{2\sigma^2(z)}\right]}\,,
\end{equation}
Essentially, for a line-of-sight halo of NFW virial mass $m$ located at redshift $z$,  $f(m,z)$ returns the PJ total mass situated on the plane of the main lens with the most similar gravitational lensing effect. Here, we do not use the mean relation reported by \citet{Despali18}, but we derive new values for the parameters (which depend on the primary lens model), for each of the lenses in our sample. The intrinsic scatter $\sigma(z)$ of the mass-redshift relation is also not the same as the one reported by \citet{Despali18}, but it is a sum in quadrature of the error on the observed mass and the uncertainty related to the scatter/different choice of the concentration-mass relation. The scatter $\sigma(z)$ does not account therefore for the measurement error on the main lens parameters. We have found this to be smaller than the scatter due to the perturber redshift degeneracy and mass density profile \citep{Despali18}, once incorrect modelling of the main deflector (e.g. wrong parametrisation) has been ruled out or addressed. 
 
 
\section{Prior and posterior probabilities}
\label{sec:posterior}

The target parameters of the model $\theta$, include the slope of the substructure mass function $\alpha$ (equation \ref{eq:sub_mf}), the mean projected mass fraction in substructures with virial masses between $M_{\rm min}^{\rm NFW}$ and $M_{\rm max}^{\rm NFW}$, $ f_{\rm sub}$ (equation \ref{equ:mu_s}), the half-mode mass $M_{\rm hm}$, and the slope $\beta$ (equations \ref{eq:mass_f} and \ref{eq:sub_mf}). Prior probabilities on these parameters are chosen as follows: 
\begin{enumerate}
\item $\alpha$ is drawn from a normal prior density distribution centred on $\alpha = 1.9$ and with a standard deviation of 0.2. This is in agreement with dark matter-only and hydrodynamical numerical simulations \citep{Despali17}; 
\item for the normalization $f_{\rm sub}$ of the substructure mass function, we assume a uniform prior density distribution proportional to $f^{-0.5}$ between 0 and 0.2;
\item  we assume $\beta$ to be normally distributed around $\beta= -1.3$  \citep{lovell17} with a standard deviation of 0.1;
\item we adopt a logarithmic prior distribution between $10^6$ and $2\times10^{12}$ $M_\odot$ for the half-mode mass. 
\end{enumerate}
We derive the posterior probability for the mass function parameters $\theta$ from the likelihood function (equation \ref{eq:likelihood}) assuming the detection of perturbers from one lens system to another to be independent events. We explore the posterior parameter space within the prior volume using {\sc MultiNest} \citep{Feroz08}. Results are presented in the following section. 


\begin{figure}
\includegraphics[width=8cm]{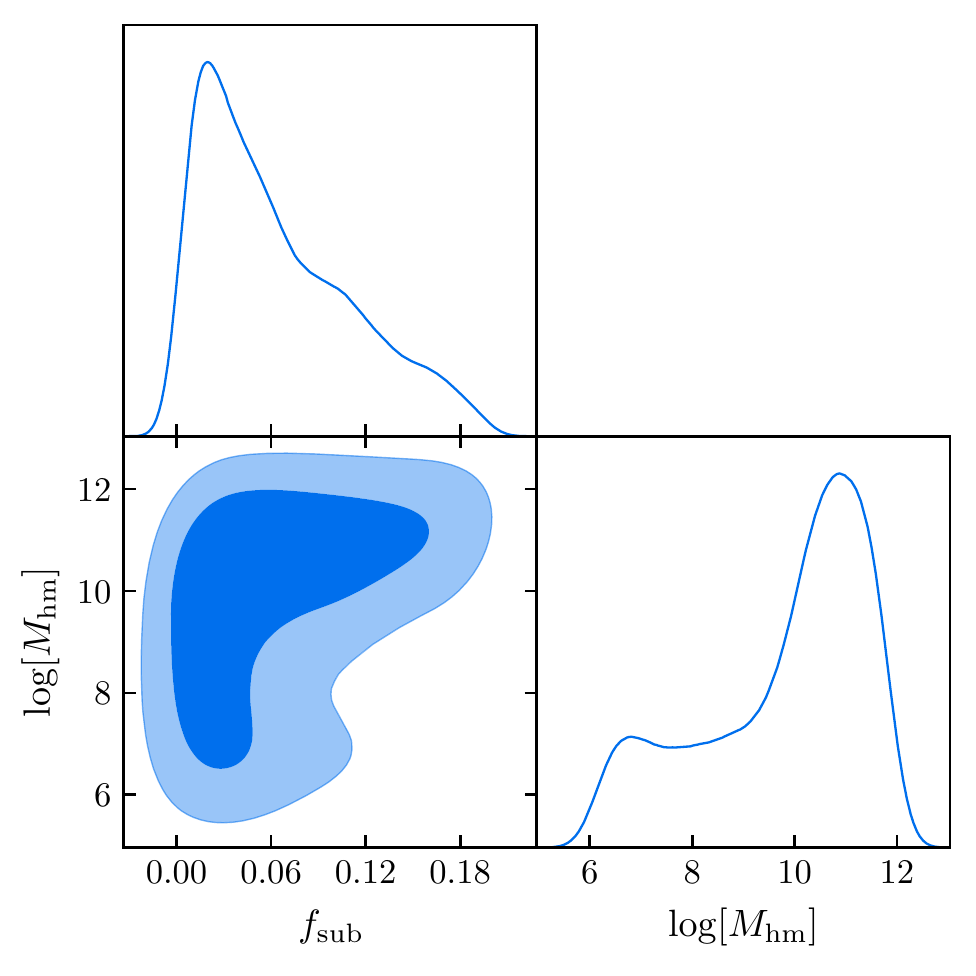}
\caption{The posterior probability distribution for the projected substructure mass fraction and the half-mode mass derived by taking into account the contribution from both substructures and line-of-sight haloes. Contours correspond to the 1$\sigma$ and 2$\sigma$ levels.} 
\label{fig:post}
\end{figure}

\begin{table}
\centering
\caption{Current and projected constraints. For the two mass function slopes, we report the mean value as well as the 68 and 95 per cent CL. For the substructure mass fraction, we report the 68 and 95 per cent upper limits, while for the half-mode mass we also report the corresponding lower limits. The first table shows the current constraints derived assuming a 10$\sigma$ detection threshold for the calculation of the sensitivity function. The second (third) table shows the projected results that one could obtain with a sample of lenses with a sensitivity which is ten (one-hundred) times better than the current one, and a single detection with the same mass and position as in the original data.}
\begin{tabular}{ccccc}
\hline
Run&Parameter & mean & $\sigma_{68}$ &  $\sigma_{95}$\\
\hline
$M^{\rm PJ}_{\rm low}$\\
\hline
&$\alpha$&1.87&-0.20~|~+0.18&-0.33~|~+0.35\\
\\
&$\beta$&-1.31&-0.09~|~+0.09&-0.17~|~+0.17\\
\\
&$f_{\rm sub}$&-&$<$ 0.087&$<$ 0.16\\
\\
&$\log M_{\rm hm}\left[M_\odot\right]$&-&9.14~|~11.9&6.42~|~12.0\\
\hline
$M^{\rm PJ}_{\rm low}/10$\\
\hline
&$\alpha$&1.81&-0.17~|~+0.14&-0.28~|~+0.31\\
\\
&$\beta$&-1.31&-0.09~|~+0.09&-0.16~|~+0.17\\
\\
&$f_{\rm sub}$&-&$< 0.089$&$< 0.17$\\
\\
&$\log M_{\rm hm}\left[M_\odot\right]$&-&10.6~|~12.1&9.40~|~12.4\\
\hline
$M^{\rm PJ}_{\rm low}/100$\\
\hline
&$\alpha$&1.84&-0.17~|~+0.17&-0.31~|~+0.33\\
\\
&$\beta$&1.31&-0.09~|~+0.09&-0.16~|~+0.17\\
\\
&$f_{\rm sub}$&-&$< 0.074$&$< 0.14$\\
\\
&$\log M_{\rm hm}\left[M_\odot\right]$&-&10.6~|~12.1&9.57~|~12.4\\
\hline
\end{tabular}
\label{tab:results}
\end{table}

\section{Results} 
\label{sec:results}

\begin{figure*}
\includegraphics[width=8.2cm]{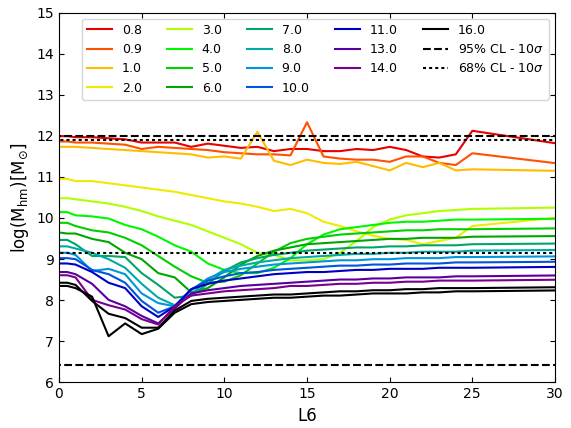}
\includegraphics[width=8cm]{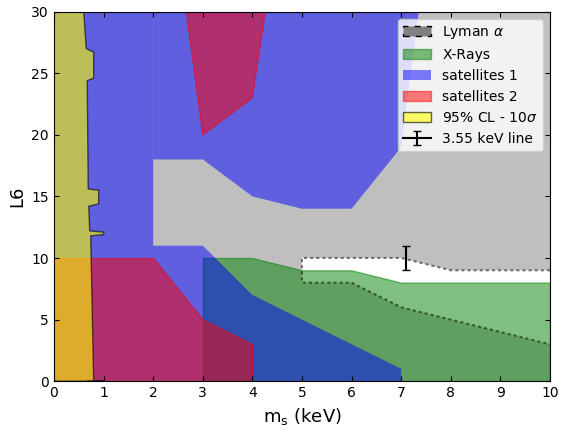}
\caption{Current constraints - Left: half-mode mass versus lepton asymmetry for different values of the neutrino mass (coloured lines), 95 and 68 per cent upper and lower limits on the half-mode mass for the current sensitivity function  (dashed and dotted black lines). Right: 95 per cent exclusion region in the $L_{\rm 6}-m_{\rm s}$ plane. The green shaded region is excluded from a non-observations of X-ray decay from M31 \citep{Watson12,Horiuchi14}. The purple and blue regions are excluded by the observed number of Milky Way satellites for two different feedback models from \citet{Lovell16}. The grey shaded region is in strong tension with Lyman $\alpha$ flux observations as described in the text. The yellow shaded region is excluded by the number of observed and non-observed mass perturbers in the sample of gravitational lens systems considered in this paper. We mark the position of the sterile neutrino model that explains the 3.55~keV line with an error bar.}
\label{fig:ms_l6}
\end{figure*}

We summarize our constraints on the model parameters in Table \ref{tab:results}. For the two slopes, $\alpha$ and $\beta$ we report the mean values, the 68 and the 95 per cent confidence levels (CLs). For the substructure mass fraction, we only report the 68 and 95 per cent upper limits, the posterior distribution being skewed towards high values. Constraints on the half-mode mass are expressed in terms of 68 and 95 per cent lower and upper limits. The posterior probability distributions for $f_{\rm sub}$ and $M_{\rm hm}$ are shown in Fig. \ref{fig:post}.

First of all, we notice that the 68 per cent upper limit on $f_{\rm sub}$ is larger than the value reported by \citet{Vegetti14} for the same sample of lenses. This difference can be attributed to a different definition of the substructure mass and mass limits as well as a different shape of the substructure mass function, which introduces a degeneracy between $f_{\rm sub}$ and $M_{\rm hm}$.

At the 95 per cent CL, we constrain the half-mode mass to be $6.42 < \log M_{\rm{hm}} \left[M_\odot\right]  < 12.0$. These limits, although rather weak, are independent of the subhalo mass function parametrization, expressed by equations (\ref{eq:mass_f}) and (\ref{eq:mass_f2}). Sterile neutrinos are a two-parameter dark matter model, where combinations of neutrino masses and lepton asymmetry in the early Universe determine the particle momentum distribution and its colder or warmer behaviour. In the left-hand panel of Fig. \ref{fig:ms_l6}, we plot the theoretical half-mode mass for different values of the neutrino mass and lepton asymmetry. Our 2$\sigma$ upper limit excludes sterile neutrino masses $m_{\rm s} < 0.8$ keV at any value of $L_{\rm 6}$. We have also derived a relationship between the mass of a thermal relic particle and the half-mode mass using the results of \citet{Viel05}, which leads to a lower limit of $m_{\rm th} > 0.3$ keV at the 2$\sigma$ level. 

In the right-hand panel of Fig. \ref{fig:ms_l6} we compare our constraints with those derived from the observed satellites in the Milky Way \citep{Lovell16}, X-ray decay searches from M31 \citep{Watson12,Horiuchi14} and Lyman $\alpha$ forest constraints. The latter measure the 1D matter power spectrum of Lyman $\alpha$ flux in QSO spectra. Comparing the limits from these studies with our results is complicated because their constraints are calculated using thermal relic matter power spectra, and a proper analysis requires simulations of structure formation to model the non-linear evolution of the power spectrum such as the flow from large scales to small scales. Another uncertainty is the thermal history around $z\sim5$, where \citet{Irsic17a} find their preferred power-law prior requires $m_{\rm th}>5.3$~keV at the 95~per~cent CL, whereas a freer prior on the thermal history relaxes the bound to $>3.5$~keV at 95~per~cent CL.

We therefore take the following approach. We draw an exclusion region based on all sterile neutrino models that have a 1D power spectrum with less power at any point in the wavenumber range $1<k<10h/\rmn{Mpc}$ than the 3.5~keV thermal relic, where $<10h/\rmn{Mpc}$ is the range of wavenumbers used in the analysis of \citet{Irsic17a}. When combined with the X-ray limit, this limit rules out all but a sliver of parameter space, which lies in the range $m_{\rm s}>5$~keV, $L_6\sim[8-10]$; the less conservative $5.3$~keV limit instead rules out all $m_{\rm s}<10$~keV. Finally, we note that our method may rule out models in which the power transfer from large scales to small scales is stronger than for the thermal relic, therefore dedicated simulations of these sterile neutrino models will be required to confirm or correct this simple model \citep[see also][]{Baur16}. 

Our 95 per cent CL exclusion regions are significantly smaller than those derived from both the satellite counts and the Lyman $\alpha$ forest, and would potentially be weaker still if the shallower slopes of sterile neutrino power spectra were taken into account fully. However, they are more robust than those from the Milky Way satellite counts, as they are less affected by feedback processes. 

\begin{figure}
\includegraphics[width=8cm]{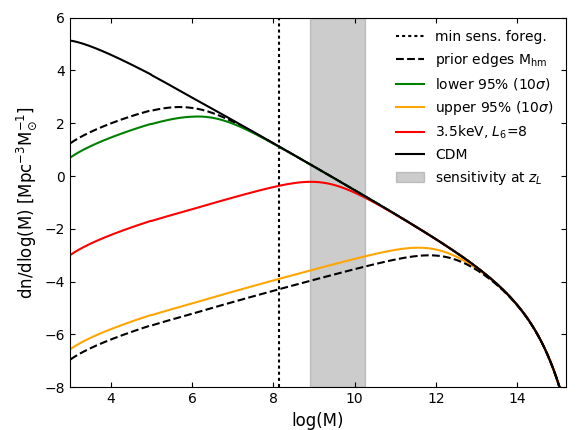}
\caption{Line-of-sight differential mass functions according to the CDM (black line) model, the sterile neutrino model compatible with the detection of the 3.55 keV line (red), and the inferred 2$-\sigma$ upper (orange) and lower (green) limits. The grey area represents the region in substructure mass probed by the current detection threshold, and the vertical dotted line the lowest detectable mass for a foreground line-of-sight halo.} 
\label{fig:mass_function}
\end{figure}

\begin{figure}
\includegraphics[width=9cm]{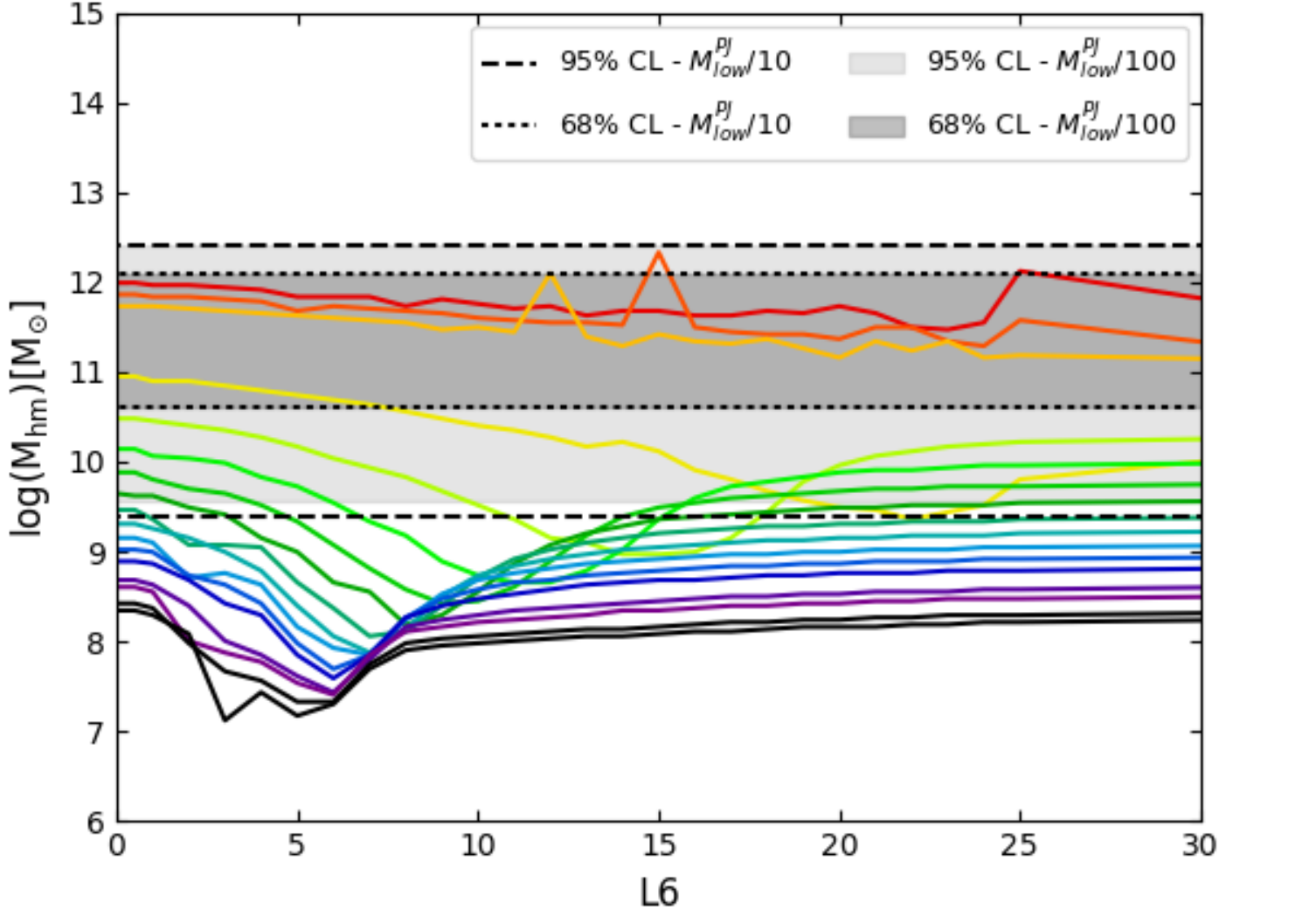}
\caption{Projected constraints - Half-mode mass versus lepton asymmetry for different values of the neutrino mass (coloured lines), 95 and 68 per cent upper and lower limits on the half-mode mass for a sensitivity function artificially improved by one and two orders of magnitudes (dashed and dotted black lines, and grey bands). We have derived these constraints under the assumption of one detected perturber with the same mass and position of the one included in the original sample.} 
\label{fig:ms_l6_2}
\end{figure}

We note that the derived 95 per cent lower limit on $M_{\rm{hm}}$ is slightly larger than the lower limit imposed by our prior. However, this constraint is still prior dominated. In Fig. \ref{fig:mass_function}, we plot the differential line-of-sight mass function corresponding to the CDM model, the sterile neutrino model compatible with the 3.5 keV emission line, and the upper and lower limits derived in this paper. Within the mass limits imposed by the sensitivity of the data (expressed now as virial NFW mass), the CDM mass function (and, therefore, any model colder than $\log M_{\rm{hm}}=6.0$) is virtually indistinguishable from the one corresponding to our lower limit. We can conclude, therefore, that our current results are not in tension with the prediction from CDM. Indeed the expected number of detectable line-of-sight haloes is 0.8$\pm$0.9, in agreement with the single detection considered in this paper. In particular, we find that a set of data with a better sensitivity will be required to constrain models with $\log M_{\rm{hm}}<6.0$. For example, as shown in Fig. \ref{fig:ms_l6_2} and Table \ref{tab:results}, the same small sample of eleven lenses considered in this paper but with a sensitivity improved by one or two orders of magnitudes, would result in a posterior probability distribution for the half-mode mass which is significantly shifted towards larger values, and more stringent constraints on both limits of the half-mode mass. Under the assumption that only one perturber, with the same mass as the one reported by \citet{Vegetti10}, is detected, such a sample of lenses would allow us to rule out CDM at the 2$\sigma$ level.

\section{Conclusions} 
\label{sec:conc}

Sterile neutrinos with masses of a few keV have been shown to be a viable candidate for dark matter particles. Depending on the level of lepton asymmetry in the early Universe, these can behave as warmer or cooler models and introduce a cut-off in the primordial linear matter power spectrum. As a consequence, structure formation within sterile neutrino cosmologies is suppressed at a scale that is a function of the neutrino mass and the lepton asymmetry.
In particular, 7 keV sterile neutrinos have been shown to be a possible explanation for the apparent 3.5 keV X-ray line detection in several galaxy clusters as well as M31 and at the centre of the Milky Way.
Recently, \citet{Lovell16} have used semi-analytical models with different sterile neutrino models to quantify the predicted number of luminous satellites in the Milky Way halo and concluded that the 7 keV sterile neutrinos are in good agreement with current observations. However, constraints on dark matter from the Milky Way satellites are limited by our knowledge of the Milky Way halo mass and the details of feedback processes. 

Strong gravitational lensing allows one to gain new insight into the properties of dark matter by quantifying the number of substructure in gravitational lens galaxies and small-mass haloes along their line-of-sight, in a way that is less affected by galaxy formation models.  
In this paper, we have used the non-detections of mass perturbers obtained by \citet{Vegetti14} in combination with the detection in the lens system SDSS J0946+1006  by \citet{Vegetti10} to derive constraints on the two-parameter model of sterile neutrinos. At the 2$-\sigma$ level we have excluded models with $m_{\rm s} < 0.8$ keV at any value of $L_{\rm 6}$.

These constraints are currently less stringent than those provided by other, more established methods. We have made explicit comparisons to the Milky Way satellite counts \citep{Polisensky11,Lovell14,Kennedy14,Lovell16,Cherry17}, which almost always prefer a thermal relic mass $>$2~keV and, in the case of \citet{Lovell16}, set constraints on sterile neutrinos of mass $<$7~keV when applying their fiducial galaxy formation model;  but their models with weaker reionization feedback can, however, evade the constraints. There are also uncertainties in the abundance and spatial distribution of the observed Milky Way satellites, in the halo mass and, the issue that this analysis is limited to the Milky Way and Andromeda systems.

Still stronger constraints are obtained from the analysis of the matter power spectrum of perturbations in the Lyman $\alpha$ forest \citep{Schneider16}. In particular, high redshift QSO spectra from XQ-100, HIRES and MIKE have been argued to rule out all values of $L_6$ for sterile neutrino mass $<$10~keV; however, different thermal histories for the intergalactic medium return weaker limits \citep{Baur17}. X-ray constraints instead set an upper limit on the sterile neutrino mass and a corresponding lower limit on $L_{6}$. The strongest limits, particularly for the 3.55~keV line, have arguably been obtained using deep XMM-Newton observations of the Draco dwarf spheroidal galaxy, but the long integration times required for a detection, coupled to uncertainties in how best to conduct those observations, reduce the possibility that further studies can improve the situation in the near future \citep[c.f.][]{Jeltema16,Ruchayskiy16}. Compared to all of the above methods, lensing studies benefit from the fact that they are sensitive to the total perturber mass and they are less affected by baryonic processes and should be therefore more robust. However, our current constraints are valid as long as the mass function parametrisation adopted here is valid for any value of $L_6$ and $m_{\rm s}$. We plan to test this hypothesis with numerical simulations of sterile neutrino models that span the range in $L_6$ inferred from the 3.5~keV line in a follow-up paper.

In agreement with \citet{Despali18}, we have found that including the contribution of line-of-sight haloes can provide a boost to the constraining power of gravitational lensing. However, due to the low redshift of the current sources and lenses, this effect is relatively small. In the near future, we will provide more stringent constraints on sterile neutrino models and the properties of dark matter in general, by using higher resolution data and samples of gravitational lens systems with a higher combination of source and lens redshift.


\appendix
\section{The Likelihood function}
\label{sec:likel}

To quantify the number of detections and non-detections of perturbers we consider a total number of $I$ bins in mass and $J$ bins in position. The $i$-th bin in mass corresponds to masses in the range $[m_i^{\rm o} , m_i^{\rm o} + \Delta m^{\rm o}]$, while the $j$-th bin in position corresponds to projected positions in the range of $[x_j^{\rm o} , x_j ^{\rm o}+ \Delta x^{\rm o}]$ \footnote{For simplicity we have reduced the number of position dimensions to just one. None the less the derivation can be analogously done in two dimensions.}. These bins also quantify the non-detections as the number density of perturbers in each bin $n_{ji}$ can be equal to zero. 

Here, we derive the Likelihood of detecting $\{n_{ij}\}_{i=1}^{I}~_{j=1}^{J}$ objects. As we are interested in including the contribution from both subhaloes and line-of-sight haloes, $n_{ij}$ is given by the sum of the two populations in each bin $ij$, i.e. $n_{ij}= n_{s,ij}+ n_{l,ij}$. We assume $n_{s,ij}$ and $n_{l,ij}$ to be distributed according to a Poisson probability density with a mean $\mu_{s,ij}$ and $\mu_{l,ij}$, respectively. Taking advantage of the fact that the sum of two Poisson distributed numbers is also Poisson distributed with a mean given by the sum of the two individual means we can write
\begin{equation}
 P\left(n_{ij}  | \mu_{ij}\right)  = \frac{  e^{- \mu_{ij}} \mu_{ij}^{n_{ij} } }{n_{ij} !}\,,
\end{equation}
with
\begin{equation}
\mu_{ij} = { \mu}_{s,ij}  + {\mu}_{l,ij}\,. 
\end{equation}
For a given set of the mass function parameters $\theta$, different $n_{ij}$ are conditionally independent of each other, so that 
\begin{multline}
 \log \left ( P\left(\{n_{ij}\}_{i=1}^{I}~_{j=1}^{J} | \theta\right)  \right) \\ = \log \left( \prod_{i=1}^{I}\prod_{j=1}^{J} P\left(n_{ij}  | \theta\right) \right) = \sum_{i=1}^{I}\sum_{j=1}^{J} \log \left( P\left(n_{ij}  | \theta\right) \right)  
\\
=  - \sum_{i=1}^{I}\sum_{j=1}^{J}   \mu_{ij}  +   \sum_{i=1}^{I}\sum_{j=1}^{J}  \log \left( \frac{\mu_{ij}^{n_{ij} } }{n_{ij} !}\right)\,.
\label{eq:logl}
\end{multline}
Under the assumptions that the bin widths are small, we can re-write the first term in the above equation as an integral of the predicted mean number density of subhaloes and line-of-sight haloes (see Section \ref{sec:exp_val}) over the masses and positions,
\begin{multline}
\sum_{i=1}^{I}\sum_{j=1}^{J}   \mu_{ij} =  \sum_{i=1}^{I}\sum_{j=1}^{J} \left[ { \mu}_{s,ij}   + {\mu}_{l,ij}  \right] \\
 =  \sum_{i=1}^{I}\sum_{j=1}^{J} \left[ \mu_{s}\left(m^{\rm o}_i,x^{\rm o}_j\right)   + \mu_{l}\left(m^{\rm o}_i,x^{\rm o}_j\right)  \right]  dm^{\rm o}  dx^{\rm o} \\
 = \int dm^{\rm o} \int dx^{\rm o}  \left[  \mu_{s}\left(m^{\rm o},x^{\rm o}\right)   + \mu_{l}\left(m^{\rm o},x^{\rm o}\right)  \right]\,.
\end{multline}
In particular, given that the number of detected objects is finite we can choose the widths of the bins small enough that the maximum number of detections per bin $ij$ is 1. 
This assumption is possible because the probability of multiple perturbers matching perfectly in mass and position is insignificantly low. We can then rewrite the second term in equation (\ref{eq:logl}) as the sum over those bins $ij$
in which a detection with $m^{\rm o}_i = m_k^{\rm ob}$ and $x^{\rm o}_j = x_k^{\rm ob}$ has been made. The detection of $n$ objects then leads to 
\begin{equation}
\sum_{k=1}^n \log \left[\mu_{s}\left(m^{\rm ob}_k,x^{\rm ob}_k\right)dm^{\rm o}  dx^{\rm o}   + \mu_{l}\left(m^{\rm ob}_k,x^{\rm ob}_k\right)  dm^{\rm o}  dx^{\rm o}\right] \,,
\end{equation}
where all other terms have vanished, since $\log ( \frac{\mu_{ij}^{0 } }{0 !} ) = 0$. 

Combining the above considerations, we can express the Likelihood function as
\begin{multline}
 \log \left ( P\left(  \{n_{ij}\}_{i=1}^{I}~_{j=1}^{J} | \theta\right)  \right) \\ =
-\int dm^{\rm o}  \int  dx^{\rm o}  \left[ \mu_{s}\left(m^{\rm o},x^{\rm o}\right)   + \mu_{l}\left(m^{\rm o},x^{\rm o}\right)  \right]  \\+ \sum_{k=1}^n  \log\left[\mu_{s}\left(m^{\rm ob}_k,x^{\rm ob}_k\right)  dm^{\rm o} dx^{\rm o}  + \mu_{l}\left(m^{\rm ob}_k,x^{\rm ob}_k\right)   dm^{\rm o} dx^{\rm o}\right] \,.
\end{multline}
For a given set of detections with $m^{\rm o}_i = m_k^{\rm ob}$ and $x^{\rm o}_j = x_k^{\rm ob}$ the number of objects in each bin $n_{ij}$ is determined, hence we can write $P\left(\{n_{ij}\}_{i=1}^{I}~_{j=1}^{J} | \theta\right)$
as  a $P\left(\left\{m^{\rm ob}_1, ....,m^{\rm ob}_n\right\},\left\{{\bf x}^{\rm ob}_1, ....,{\bf x}^{\rm ob}_n\right\}| \theta\right)$. 

\section*{Acknowledgements}
The authors are grateful to A. Boyarsky, O. Ruchayskiy and S.~D.~M. White for useful comments and discussions.
M.~R.~L. is  supported  by  a COFUND/Durham  Junior Research Fellowship under EU grant 609412.
M.~R.~L. acknowledges support by a Grant of Excellence from the Icelandic Research Fund (grant number 173929$-$051).
S.~V. has received funding from the European Research Council (ERC) under the Europen Union's Horizon 2020 research and innovation programme (grant agreement No 758853).

\bibliography{ms}

\begin{thebibliography}{}
\makeatletter
\relax
\def\mn@urlcharsother{\let\do\@makeother \do\$\do\&\do\#\do\^\do\_\do\%\do\~}
\def\mn@doi{\begingroup\mn@urlcharsother \@ifnextchar [ {\mn@doi@}
  {\mn@doi@[]}}
\def\mn@doi@[#1]#2{\def\@tempa{#1}\ifx\@tempa\@empty \href
  {http://dx.doi.org/#2} {doi:#2}\else \href {http://dx.doi.org/#2} {#1}\fi
  \endgroup}
\def\mn@eprint#1#2{\mn@eprint@#1:#2::\@nil}
\def\mn@eprint@arXiv#1{\href {http://arxiv.org/abs/#1} {{\tt arXiv:#1}}}
\def\mn@eprint@dblp#1{\href {http://dblp.uni-trier.de/rec/bibtex/#1.xml}
  {dblp:#1}}
\def\mn@eprint@#1:#2:#3:#4\@nil{\def\@tempa {#1}\def\@tempb {#2}\def\@tempc
  {#3}\ifx \@tempc \@empty \let \@tempc \@tempb \let \@tempb \@tempa \fi \ifx
  \@tempb \@empty \def\@tempb {arXiv}\fi \@ifundefined
  {mn@eprint@\@tempb}{\@tempb:\@tempc}{\expandafter \expandafter \csname
  mn@eprint@\@tempb\endcsname \expandafter{\@tempc}}}

\bibitem[\protect\citeauthoryear{{Aharonian} et~al.,}{{Aharonian}
  et~al.}{2017}]{Aharonian16}
{Aharonian} F.~A.,  et~al., 2017, \mn@doi [\apjl] {10.3847/2041-8213/aa61fa},
  \href {http://adsabs.harvard.edu/abs/2017ApJ...837L..15A} {837, L15}

\bibitem[\protect\citeauthoryear{{Amorisco} \& {Evans}}{{Amorisco} \&
  {Evans}}{2012}]{Amorisco2012}
{Amorisco} N.~C.,  {Evans} N.~W.,  2012, \mn@doi [\mnras]
  {10.1111/j.1365-2966.2011.19684.x}, \href
  {http://adsabs.harvard.edu/abs/2012MNRAS.419..184A} {419, 184}

\bibitem[\protect\citeauthoryear{{Anderson}, {Churazov}  \&
  {Bregman}}{{Anderson} et~al.}{2015}]{Anderson14}
{Anderson} M.~E.,  {Churazov} E.,   {Bregman} J.~N.,  2015, \mn@doi [\mnras]
  {10.1093/mnras/stv1559}, \href
  {http://adsabs.harvard.edu/abs/2015MNRAS.452.3905A} {452, 3905}

\bibitem[\protect\citeauthoryear{{Asaka} \& {Shaposhnikov}}{{Asaka} \&
  {Shaposhnikov}}{2005}]{Asaka05}
{Asaka} T.,  {Shaposhnikov} M.,  2005, \mn@doi [Physics Letters B]
  {10.1016/j.physletb.2005.06.020}, \href
  {http://adsabs.harvard.edu/abs/2005PhLB..620...17A} {620, 17}

\bibitem[\protect\citeauthoryear{{Auger}, {Treu}, {Bolton}, {Gavazzi},
  {Koopmans}, {Marshall}, {Moustakas}  \& {Burles}}{{Auger}
  et~al.}{2010}]{Auger10}
{Auger} M.~W.,  {Treu} T.,  {Bolton} A.~S.,  {Gavazzi} R.,  {Koopmans}
  L.~V.~E.,  {Marshall} P.~J.,  {Moustakas} L.~A.,   {Burles} S.,  2010,
  \mn@doi [\apj] {10.1088/0004-637X/724/1/511}, \href
  {http://adsabs.harvard.edu/abs/2010ApJ...724..511A} {724, 511}

\bibitem[\protect\citeauthoryear{{Barnab{\`e}}, {Nipoti}, {Koopmans}, {Vegetti}
   \& {Ciotti}}{{Barnab{\`e}} et~al.}{2009}]{Barnabe09}
{Barnab{\`e}} M.,  {Nipoti} C.,  {Koopmans} L.~V.~E.,  {Vegetti} S.,   {Ciotti}
  L.,  2009, \mn@doi [\mnras] {10.1111/j.1365-2966.2008.14208.x}, \href
  {http://adsabs.harvard.edu/abs/2009MNRAS.393.1114B} {393, 1114}

\bibitem[\protect\citeauthoryear{{Baur}, {Palanque-Delabrouille}, {Y{\`e}che},
  {Magneville}  \& {Viel}}{{Baur} et~al.}{2016}]{Baur16}
{Baur} J.,  {Palanque-Delabrouille} N.,  {Y{\`e}che} C.,  {Magneville} C.,
  {Viel} M.,  2016, \mn@doi [\jcap] {10.1088/1475-7516/2016/08/012}, \href
  {http://adsabs.harvard.edu/abs/2016JCAP...08..012B} {8, 012}

\bibitem[\protect\citeauthoryear{{Baur}, {Palanque-Delabrouille}, {Y{\`e}che},
  {Boyarsky}, {Ruchayskiy}, {Armengaud}  \& {Lesgourgues}}{{Baur}
  et~al.}{2017}]{Baur17}
{Baur} J.,  {Palanque-Delabrouille} N.,  {Y{\`e}che} C.,  {Boyarsky} A.,
  {Ruchayskiy} O.,  {Armengaud} {\'E}.,   {Lesgourgues} J.,  2017, \mn@doi
  [\jcap] {10.1088/1475-7516/2017/12/013}, \href
  {http://adsabs.harvard.edu/abs/2017JCAP...12..013B} {12, 013}

\bibitem[\protect\citeauthoryear{{Birrer}, {Amara}  \& {Refregier}}{{Birrer}
  et~al.}{2017}]{Birrer17}
{Birrer} S.,  {Amara} A.,   {Refregier} A.,  2017, \mn@doi [\jcap]
  {10.1088/1475-7516/2017/05/037}, \href
  {http://adsabs.harvard.edu/abs/2017JCAP...05..037B} {5, 037}

\bibitem[\protect\citeauthoryear{{Bolton}, {Burles}, {Koopmans}, {Treu}  \&
  {Moustakas}}{{Bolton} et~al.}{2006}]{Bolton06}
{Bolton} A.~S.,  {Burles} S.,  {Koopmans} L.~V.~E.,  {Treu} T.,   {Moustakas}
  L.~A.,  2006, \mn@doi [\apj] {10.1086/498884}, \href
  {http://adsabs.harvard.edu/abs/2006ApJ...638..703B} {638, 703}

\bibitem[\protect\citeauthoryear{{Bottino}, {Fornengo}  \& {Scopel}}{{Bottino}
  et~al.}{2003}]{Bottino03}
{Bottino} A.,  {Fornengo} N.,   {Scopel} S.,  2003, \mn@doi [\prd]
  {10.1103/PhysRevD.67.063519}, \href
  {http://adsabs.harvard.edu/abs/2003PhRvD..67f3519B} {67, 063519}

\bibitem[\protect\citeauthoryear{{Boyarsky}, {Ruchayskiy}  \&
  {Shaposhnikov}}{{Boyarsky} et~al.}{2009}]{Boyarsky09a}
{Boyarsky} A.,  {Ruchayskiy} O.,   {Shaposhnikov} M.,  2009, \mn@doi [Annual
  Review of Nuclear and Particle Science] {10.1146/annurev.nucl.010909.083654},
  \href {http://adsabs.harvard.edu/abs/2009ARNPS..59..191B} {59, 191}

\bibitem[\protect\citeauthoryear{{Boyarsky}, {Iakubovskyi}  \&
  {Ruchayskiy}}{{Boyarsky} et~al.}{2012}]{Boyarsky12}
{Boyarsky} A.,  {Iakubovskyi} D.,   {Ruchayskiy} O.,  2012, \mn@doi [Physics of
  the Dark Universe] {10.1016/j.dark.2012.11.001}, \href
  {http://adsabs.harvard.edu/abs/2012PDU.....1..136B} {1, 136}

\bibitem[\protect\citeauthoryear{{Boyarsky}, {Ruchayskiy}, {Iakubovskyi}  \&
  {Franse}}{{Boyarsky} et~al.}{2014}]{Boyarsky14}
{Boyarsky} A.,  {Ruchayskiy} O.,  {Iakubovskyi} D.,   {Franse} J.,  2014,
  \mn@doi [Physical Review Letters] {10.1103/PhysRevLett.113.251301}, \href
  {http://adsabs.harvard.edu/abs/2014PhRvL.113y1301B} {113, 251301}

\bibitem[\protect\citeauthoryear{Boyarsky, Franse, Iakubovskyi  \&
  Ruchayskiy}{Boyarsky et~al.}{2015}]{Boyarsky15}
Boyarsky A.,  Franse J.,  Iakubovskyi D.,   Ruchayskiy O.,  2015, \mn@doi
  [Phys. Rev. Lett.] {10.1103/PhysRevLett.115.161301}, 115, 161301

\bibitem[\protect\citeauthoryear{{Boylan-Kolchin}, {Bullock}  \&
  {Kaplinghat}}{{Boylan-Kolchin} et~al.}{2012}]{Boylan2012}
{Boylan-Kolchin} M.,  {Bullock} J.~S.,   {Kaplinghat} M.,  2012, \mn@doi
  [\mnras] {10.1111/j.1365-2966.2012.20695.x}, \href
  {http://adsabs.harvard.edu/abs/2012MNRAS.422.1203B} {422, 1203}

\bibitem[\protect\citeauthoryear{{Bulbul}, {Markevitch}, {Foster}, {Smith},
  {Loewenstein}  \& {Randall}}{{Bulbul} et~al.}{2014}]{Bulbul14}
{Bulbul} E.,  {Markevitch} M.,  {Foster} A.,  {Smith} R.~K.,  {Loewenstein} M.,
    {Randall} S.~W.,  2014, \mn@doi [\apj] {10.1088/0004-637X/789/1/13}, \href
  {http://adsabs.harvard.edu/abs/2014ApJ...789...13B} {789, 13}

\bibitem[\protect\citeauthoryear{{Cappelluti} et~al.,}{{Cappelluti}
  et~al.}{2017}]{Cappelluti17}
{Cappelluti} N.,  et~al., 2017, preprint, \href
  {http://adsabs.harvard.edu/abs/2017arXiv170107932C} {} (\mn@eprint {arXiv}
  {1701.07932})

\bibitem[\protect\citeauthoryear{Chatterjee \& Koopmans}{Chatterjee \&
  Koopmans}{2018}]{Chatterjee17}
Chatterjee S.,  Koopmans L. V.~E.,  2018, \mn@doi [Monthly Notices of the Royal
  Astronomical Society] {10.1093/mnras/stx2674}, 474, 1762

\bibitem[\protect\citeauthoryear{{Cherry} \& {Horiuchi}}{{Cherry} \&
  {Horiuchi}}{2017}]{Cherry17}
{Cherry} J.~F.,  {Horiuchi} S.,  2017, \mn@doi [\prd]
  {10.1103/PhysRevD.95.083015}, \href
  {http://adsabs.harvard.edu/abs/2017PhRvD..95h3015C} {95, 083015}

\bibitem[\protect\citeauthoryear{{Cyr-Racine}, {Moustakas}, {Keeton},
  {Sigurdson}  \& {Gilman}}{{Cyr-Racine} et~al.}{2016}]{Cyr-Racine16}
{Cyr-Racine} F.-Y.,  {Moustakas} L.~A.,  {Keeton} C.~R.,  {Sigurdson} K.,
  {Gilman} D.~A.,  2016, \mn@doi [\prd] {10.1103/PhysRevD.94.043505}, \href
  {http://adsabs.harvard.edu/abs/2016PhRvD..94d3505C} {94, 043505}

\bibitem[\protect\citeauthoryear{{Dalal} \& {Kochanek}}{{Dalal} \&
  {Kochanek}}{2002}]{Dalal02}
{Dalal} N.,  {Kochanek} C.~S.,  2002, \mn@doi [\apj] {10.1086/340303}, \href
  {http://adsabs.harvard.edu/abs/2002ApJ...572...25D} {572, 25}

\bibitem[\protect\citeauthoryear{{Daylan}, {Cyr-Racine}, {Diaz Rivero},
  {Dvorkin}  \& {Finkbeiner}}{{Daylan} et~al.}{2017}]{Daylan17}
{Daylan} T.,  {Cyr-Racine} F.-Y.,  {Diaz Rivero} A.,  {Dvorkin} C.,
  {Finkbeiner} D.~P.,  2017, preprint, \href
  {http://adsabs.harvard.edu/abs/2017arXiv170606111D} {} (\mn@eprint {arXiv}
  {1706.06111})

\bibitem[\protect\citeauthoryear{{Despali} \& {Vegetti}}{{Despali} \&
  {Vegetti}}{2017}]{Despali17}
{Despali} G.,  {Vegetti} S.,  2017, \mn@doi [\mnras] {10.1093/mnras/stx966},
  \href {http://adsabs.harvard.edu/abs/2017MNRAS.469.1997D} {469, 1997}

\bibitem[\protect\citeauthoryear{{Despali}, {Giocoli}, {Angulo}, {Tormen},
  {Sheth}, {Baso}  \& {Moscardini}}{{Despali} et~al.}{2016}]{Despali16}
{Despali} G.,  {Giocoli} C.,  {Angulo} R.~E.,  {Tormen} G.,  {Sheth} R.~K.,
  {Baso} G.,   {Moscardini} L.,  2016, \mn@doi [\mnras]
  {10.1093/mnras/stv2842}, \href
  {http://adsabs.harvard.edu/abs/2016MNRAS.456.2486D} {456, 2486}

\bibitem[\protect\citeauthoryear{{Despali}, {Vegetti}, {White}, {Giocoli}  \&
  {van den Bosch}}{{Despali} et~al.}{2018}]{Despali18}
{Despali} G.,  {Vegetti} S.,  {White} S.~D.~M.,  {Giocoli} C.,   {van den
  Bosch} F.~C.,  2018, \mn@doi [\mnras] {10.1093/mnras/sty159}, \href
  {http://adsabs.harvard.edu/abs/2018MNRAS.475.5424D} {475, 5424}

\bibitem[\protect\citeauthoryear{Dolgov \& Hansen}{Dolgov \&
  Hansen}{2002}]{Dolgov:00}
Dolgov A.,  Hansen S.,  2002, \mn@doi [Astropart.Phys.]
  {10.1016/S0927-6505(01)00115-3}, 16, 339

\bibitem[\protect\citeauthoryear{{Duffy}, {Schaye}, {Kay}  \& {Dalla
  Vecchia}}{{Duffy} et~al.}{2008}]{Duffy08}
{Duffy} A.~R.,  {Schaye} J.,  {Kay} S.~T.,   {Dalla Vecchia} C.,  2008, \mn@doi
  [\mnras] {10.1111/j.1745-3933.2008.00537.x}, \href
  {http://adsabs.harvard.edu/abs/2008MNRAS.390L..64D} {390, L64}

\bibitem[\protect\citeauthoryear{{Fadely} \& {Keeton}}{{Fadely} \&
  {Keeton}}{2012}]{Fadely12}
{Fadely} R.,  {Keeton} C.~R.,  2012, \mn@doi [\mnras]
  {10.1111/j.1365-2966.2011.19729.x}, \href
  {http://adsabs.harvard.edu/abs/2012MNRAS.419..936F} {419, 936}

\bibitem[\protect\citeauthoryear{{Feroz} \& {Hobson}}{{Feroz} \&
  {Hobson}}{2008}]{Feroz08}
{Feroz} F.,  {Hobson} M.~P.,  2008, \mn@doi [\mnras]
  {10.1111/j.1365-2966.2007.12353.x}, \href
  {http://adsabs.harvard.edu/abs/2008MNRAS.384..449F} {384, 449}

\bibitem[\protect\citeauthoryear{{Garzilli}, {Boyarsky}  \&
  {Ruchayskiy}}{{Garzilli} et~al.}{2017}]{Garzilli17}
{Garzilli} A.,  {Boyarsky} A.,   {Ruchayskiy} O.,  2017, \mn@doi [Physics
  Letters B] {10.1016/j.physletb.2017.08.022}, \href
  {http://adsabs.harvard.edu/abs/2017PhLB..773..258G} {773, 258}

\bibitem[\protect\citeauthoryear{{Gilman}, {Agnello}, {Treu}, {Keeton}  \&
  {Nierenberg}}{{Gilman} et~al.}{2017}]{Gilman17}
{Gilman} D.,  {Agnello} A.,  {Treu} T.,  {Keeton} C.~R.,   {Nierenberg} A.~M.,
  2017, \mn@doi [\mnras] {10.1093/mnras/stx158}, \href
  {http://adsabs.harvard.edu/abs/2017MNRAS.467.3970G} {467, 3970}

\bibitem[\protect\citeauthoryear{{Gilman}, {Birrer}, {Treu}  \&
  {Keeton}}{{Gilman} et~al.}{2018}]{Gilman18}
{Gilman} D.,  {Birrer} S.,  {Treu} T.,   {Keeton} C.~R.,  2018, preprint, \href
  {http://adsabs.harvard.edu/abs/2017arXiv171204945G} {} (\mn@eprint {arXiv}
  {1712.04945})

\bibitem[\protect\citeauthoryear{{Gu}, {Kaastra}, {Raassen}, {Mullen},
  {Cumbee}, {Lyons}  \& {Stancil}}{{Gu} et~al.}{2015}]{Gu15}
{Gu} L.,  {Kaastra} J.,  {Raassen} A.~J.~J.,  {Mullen} P.~D.,  {Cumbee} R.~S.,
  {Lyons} D.,   {Stancil} P.~C.,  2015, \mn@doi [\aap]
  {10.1051/0004-6361/201527634}, \href
  {http://adsabs.harvard.edu/abs/2015A%26A...584L..11G} {584, L11}

\bibitem[\protect\citeauthoryear{{Hezaveh} et~al.,}{{Hezaveh}
  et~al.}{2016}]{Hezaveh16}
{Hezaveh} Y.~D.,  et~al., 2016, \mn@doi [\apj] {10.3847/0004-637X/823/1/37},
  \href {http://adsabs.harvard.edu/abs/2016ApJ...823...37H} {823, 37}

\bibitem[\protect\citeauthoryear{{Horiuchi}, {Humphrey}, {O{\~n}orbe},
  {Abazajian}, {Kaplinghat}  \& {Garrison-Kimmel}}{{Horiuchi}
  et~al.}{2014}]{Horiuchi14}
{Horiuchi} S.,  {Humphrey} P.~J.,  {O{\~n}orbe} J.,  {Abazajian} K.~N.,
  {Kaplinghat} M.,   {Garrison-Kimmel} S.,  2014, \mn@doi [\prd]
  {10.1103/PhysRevD.89.025017}, \href
  {http://adsabs.harvard.edu/abs/2014PhRvD..89b5017H} {89, 025017}

\bibitem[\protect\citeauthoryear{{Hsueh}, {Fassnacht}, {Vegetti}, {McKean},
  {Spingola}, {Auger}, {Koopmans}  \& {Lagattuta}}{{Hsueh}
  et~al.}{2016}]{Hsueh16}
{Hsueh} J.-W.,  {Fassnacht} C.~D.,  {Vegetti} S.,  {McKean} J.~P.,  {Spingola}
  C.,  {Auger} M.~W.,  {Koopmans} L.~V.~E.,   {Lagattuta} D.~J.,  2016, \mn@doi
  [\mnras] {10.1093/mnrasl/slw146}, \href
  {http://adsabs.harvard.edu/abs/2016MNRAS.463L..51H} {463, L51}

\bibitem[\protect\citeauthoryear{{Hsueh}, {Despali}, {Vegetti}, {Xu},
  {Fassnacht}  \& {Metcalf}}{{Hsueh} et~al.}{2017a}]{Hsueh17b}
{Hsueh} J.-W.,  {Despali} G.,  {Vegetti} S.,  {Xu} D.,  {Fassnacht} C.~D.,
  {Metcalf} R.~B.,  2017a, preprint, \href
  {http://adsabs.harvard.edu/abs/2017arXiv170707680H} {} (\mn@eprint {arXiv}
  {1707.07680})

\bibitem[\protect\citeauthoryear{{Hsueh} et~al.,}{{Hsueh}
  et~al.}{2017b}]{Hsueh17a}
{Hsueh} J.-W.,  et~al., 2017b, \mn@doi [\mnras] {10.1093/mnras/stx1082}, \href
  {http://adsabs.harvard.edu/abs/2017MNRAS.469.3713H} {469, 3713}

\bibitem[\protect\citeauthoryear{{Iakubovskyi}}{{Iakubovskyi}}{2014}]{Iakubovskyi14}
{Iakubovskyi} D.~A.,  2014, Advances in Astronomy and Space Physics, \href
  {http://adsabs.harvard.edu/abs/2014AASP....4....9I} {4, 9}

\bibitem[\protect\citeauthoryear{{Ir{\v s}i{\v c}} et~al.,}{{Ir{\v s}i{\v c}}
  et~al.}{2017a}]{Irsic17a}
{Ir{\v s}i{\v c}} V.,  et~al., 2017a, \mn@doi [\prd]
  {10.1103/PhysRevD.96.023522}, \href
  {http://adsabs.harvard.edu/abs/2017PhRvD..96b3522I} {96, 023522}

\bibitem[\protect\citeauthoryear{{Ir{\v s}i{\v c}}, {Viel}, {Haehnelt},
  {Bolton}  \& {Becker}}{{Ir{\v s}i{\v c}} et~al.}{2017b}]{Irsic17b}
{Ir{\v s}i{\v c}} V.,  {Viel} M.,  {Haehnelt} M.~G.,  {Bolton} J.~S.,
  {Becker} G.~D.,  2017b, \mn@doi [Physical Review Letters]
  {10.1103/PhysRevLett.119.031302}, \href
  {http://adsabs.harvard.edu/abs/2017PhRvL.119c1302I} {119, 031302}

\bibitem[\protect\citeauthoryear{{Jeltema} \& {Profumo}}{{Jeltema} \&
  {Profumo}}{2016}]{Jeltema16}
{Jeltema} T.,  {Profumo} S.,  2016, \mn@doi [\mnras] {10.1093/mnras/stw578},
  \href {http://adsabs.harvard.edu/abs/2016MNRAS.458.3592J} {458, 3592}

\bibitem[\protect\citeauthoryear{{Kennedy}, {Frenk}, {Cole}  \&
  {Benson}}{{Kennedy} et~al.}{2014}]{Kennedy14}
{Kennedy} R.,  {Frenk} C.,  {Cole} S.,   {Benson} A.,  2014, \mn@doi [\mnras]
  {10.1093/mnras/stu719}, \href
  {http://adsabs.harvard.edu/abs/2014MNRAS.442.2487K} {442, 2487}

\bibitem[\protect\citeauthoryear{{Klypin}, {Kravtsov}, {Valenzuela}  \&
  {Prada}}{{Klypin} et~al.}{1999}]{Klypin1999}
{Klypin} A.,  {Kravtsov} A.~V.,  {Valenzuela} O.,   {Prada} F.,  1999, \mn@doi
  [\apj] {10.1086/307643}, \href
  {http://adsabs.harvard.edu/abs/1999ApJ...522...82K} {522, 82}

\bibitem[\protect\citeauthoryear{{Koopmans}}{{Koopmans}}{2005}]{Koopmans05}
{Koopmans} L.~V.~E.,  2005, \mn@doi [\mnras]
  {10.1111/j.1365-2966.2005.09523.x}, \href
  {http://adsabs.harvard.edu/abs/2005MNRAS.363.1136K} {363, 1136}

\bibitem[\protect\citeauthoryear{{Kuzio de Naray}, {McGaugh}  \& {de
  Blok}}{{Kuzio de Naray} et~al.}{2008}]{Kuzio2008}
{Kuzio de Naray} R.,  {McGaugh} S.~S.,   {de Blok} W.~J.~G.,  2008, \mn@doi
  [\apj] {10.1086/527543}, \href
  {http://adsabs.harvard.edu/abs/2008ApJ...676..920K} {676, 920}

\bibitem[\protect\citeauthoryear{{Laine} \& {Shaposhnikov}}{{Laine} \&
  {Shaposhnikov}}{2008}]{Laine08}
{Laine} M.,  {Shaposhnikov} M.,  2008, \mn@doi [\jcap]
  {10.1088/1475-7516/2008/06/031}, \href
  {http://adsabs.harvard.edu/abs/2008JCAP...06..031L} {6, 31}

\bibitem[\protect\citeauthoryear{{Lewis}, {Challinor}  \& {Lasenby}}{{Lewis}
  et~al.}{2000}]{Lewis00}
{Lewis} A.,  {Challinor} A.,   {Lasenby} A.,  2000, \mn@doi [\apj]
  {10.1086/309179}, \href {http://adsabs.harvard.edu/abs/2000ApJ...538..473L}
  {538, 473}

\bibitem[\protect\citeauthoryear{{Lovell}, {Frenk}, {Eke}, {Jenkins}, {Gao}  \&
  {Theuns}}{{Lovell} et~al.}{2014}]{Lovell14}
{Lovell} M.~R.,  {Frenk} C.~S.,  {Eke} V.~R.,  {Jenkins} A.,  {Gao} L.,
  {Theuns} T.,  2014, \mn@doi [\mnras] {10.1093/mnras/stt2431}, \href
  {http://adsabs.harvard.edu/abs/2014MNRAS.439..300L} {439, 300}

\bibitem[\protect\citeauthoryear{{Lovell} et~al.,}{{Lovell}
  et~al.}{2016}]{Lovell16}
{Lovell} M.~R.,  et~al., 2016, \mn@doi [\mnras] {10.1093/mnras/stw1317}, \href
  {http://adsabs.harvard.edu/abs/2016MNRAS.461...60L} {461, 60}

\bibitem[\protect\citeauthoryear{{Lovell} et~al.,}{{Lovell}
  et~al.}{2017}]{lovell17}
{Lovell} M.~R.,  et~al., 2017, \mn@doi [\mnras] {10.1093/mnras/stx654}, \href
  {http://adsabs.harvard.edu/abs/2017MNRAS.468.4285L} {468, 4285}

\bibitem[\protect\citeauthoryear{{Ludlow}, {Bose}, {Angulo}, {Wang},
  {Hellwing}, {Navarro}, {Cole}  \& {Frenk}}{{Ludlow} et~al.}{2016}]{Ludlow16}
{Ludlow} A.~D.,  {Bose} S.,  {Angulo} R.~E.,  {Wang} L.,  {Hellwing} W.~A.,
  {Navarro} J.~F.,  {Cole} S.,   {Frenk} C.~S.,  2016, \mn@doi [\mnras]
  {10.1093/mnras/stw1046}, \href
  {http://adsabs.harvard.edu/abs/2016MNRAS.460.1214L} {460, 1214}

\bibitem[\protect\citeauthoryear{{Mao} \& {Schneider}}{{Mao} \&
  {Schneider}}{1998}]{Mao98}
{Mao} S.,  {Schneider} P.,  1998, \mn@doi [\mnras]
  {10.1046/j.1365-8711.1998.01319.x}, \href
  {http://adsabs.harvard.edu/abs/1998MNRAS.295..587M} {295, 587}

\bibitem[\protect\citeauthoryear{{Marshall}, {Tananbaum}, {Avni}  \&
  {Zamorani}}{{Marshall} et~al.}{1983}]{Marshall83}
{Marshall} H.~L.,  {Tananbaum} H.,  {Avni} Y.,   {Zamorani} G.,  1983, \mn@doi
  [\apj] {10.1086/161016}, \href
  {http://adsabs.harvard.edu/abs/1983ApJ...269...35M} {269, 35}

\bibitem[\protect\citeauthoryear{{McKean} et~al.,}{{McKean}
  et~al.}{2007}]{McKean07}
{McKean} J.~P.,  et~al., 2007, \mn@doi [\mnras]
  {10.1111/j.1365-2966.2007.11744.x}, \href
  {http://adsabs.harvard.edu/abs/2007MNRAS.378..109M} {378, 109}

\bibitem[\protect\citeauthoryear{{Metcalf}}{{Metcalf}}{2005}]{metcalf05}
{Metcalf} R.~B.,  2005, \mn@doi [\apj] {10.1086/431574}, \href
  {http://adsabs.harvard.edu/abs/2005ApJ...629..673M} {629, 673}

\bibitem[\protect\citeauthoryear{{Moore}}{{Moore}}{1994}]{Moore1994}
{Moore} B.,  1994, \mn@doi [\nat] {10.1038/370629a0}, \href
  {http://adsabs.harvard.edu/abs/1994Natur.370..629M} {370, 629}

\bibitem[\protect\citeauthoryear{{Navarro}, {Frenk}  \& {White}}{{Navarro}
  et~al.}{1997}]{Navarro97}
{Navarro} J.~F.,  {Frenk} C.~S.,   {White} S.~D.~M.,  1997, \mn@doi [\apj]
  {10.1086/304888}, \href {http://adsabs.harvard.edu/abs/1997ApJ...490..493N}
  {490, 493}

\bibitem[\protect\citeauthoryear{{Neronov}, {Malyshev}  \& {Eckert}}{{Neronov}
  et~al.}{2016}]{Neronov16}
{Neronov} A.,  {Malyshev} D.,   {Eckert} D.,  2016, \mn@doi [\prd]
  {10.1103/PhysRevD.94.123504}, \href
  {http://adsabs.harvard.edu/abs/2016PhRvD..94l3504N} {94, 123504}

\bibitem[\protect\citeauthoryear{{Nierenberg} et~al.,}{{Nierenberg}
  et~al.}{2017}]{Nierenberg17}
{Nierenberg} A.~M.,  et~al., 2017, preprint, \href
  {http://adsabs.harvard.edu/abs/2017arXiv170105188N} {} (\mn@eprint {arXiv}
  {1701.05188})

\bibitem[\protect\citeauthoryear{{Polisensky} \& {Ricotti}}{{Polisensky} \&
  {Ricotti}}{2011}]{Polisensky11}
{Polisensky} E.,  {Ricotti} M.,  2011, \mn@doi [\prd]
  {10.1103/PhysRevD.83.043506}, \href
  {http://adsabs.harvard.edu/abs/2011PhRvD..83d3506P} {83, 043506}

\bibitem[\protect\citeauthoryear{{Ringwald}}{{Ringwald}}{2016}]{Ringwald16}
{Ringwald} A.,  2016, in Proceedings of the Neutrino Oscillation Workshop
  (NOW2016). p.~81 (\mn@eprint {arXiv} {1612.08933})

\bibitem[\protect\citeauthoryear{{Robles} et~al.,}{{Robles}
  et~al.}{2017}]{Robles17}
{Robles} V.~H.,  et~al., 2017, preprint, \href
  {http://adsabs.harvard.edu/abs/2017arXiv170607514R} {} (\mn@eprint {arXiv}
  {1706.07514})

\bibitem[\protect\citeauthoryear{{Ruchayskiy} et~al.,}{{Ruchayskiy}
  et~al.}{2016}]{Ruchayskiy16}
{Ruchayskiy} O.,  et~al., 2016, \mn@doi [\mnras] {10.1093/mnras/stw1026}, \href
  {http://adsabs.harvard.edu/abs/2016MNRAS.460.1390R} {460, 1390}

\bibitem[\protect\citeauthoryear{{Schneider}}{{Schneider}}{2016}]{Schneider16}
{Schneider} A.,  2016, \mn@doi [\jcap] {10.1088/1475-7516/2016/04/059}, \href
  {http://adsabs.harvard.edu/abs/2016JCAP...04..059S} {4, 059}

\bibitem[\protect\citeauthoryear{{Schneider}, {Smith}, {Macci{\`o}}  \&
  {Moore}}{{Schneider} et~al.}{2012}]{Schneider12}
{Schneider} A.,  {Smith} R.~E.,  {Macci{\`o}} A.~V.,   {Moore} B.,  2012,
  \mn@doi [\mnras] {10.1111/j.1365-2966.2012.21252.x}, \href
  {http://adsabs.harvard.edu/abs/2012MNRAS.424..684S} {424, 684}

\bibitem[\protect\citeauthoryear{{Sheth} \& {Tormen}}{{Sheth} \&
  {Tormen}}{1999}]{Sheth99b}
{Sheth} R.~K.,  {Tormen} G.,  1999, \mnras, \href
  {http://adsabs.harvard.edu/abs/1999MNRAS.308..119S} {308, 119}

\bibitem[\protect\citeauthoryear{{Shi} \& {Fuller}}{{Shi} \&
  {Fuller}}{1999}]{Shi99}
{Shi} X.,  {Fuller} G.~M.,  1999, \mn@doi [Physical Review Letters]
  {10.1103/PhysRevLett.82.2832}, \href
  {http://adsabs.harvard.edu/abs/1999PhRvL..82.2832S} {82, 2832}

\bibitem[\protect\citeauthoryear{{Vegetti} \& {Koopmans}}{{Vegetti} \&
  {Koopmans}}{2009}]{Vegetti09a}
{Vegetti} S.,  {Koopmans} L.~V.~E.,  2009, \mn@doi [\mnras]
  {10.1111/j.1365-2966.2008.14005.x}, \href
  {http://adsabs.harvard.edu/abs/2009MNRAS.392..945V} {392, 945}

\bibitem[\protect\citeauthoryear{{Vegetti}, {Czoske}  \& {Koopmans}}{{Vegetti}
  et~al.}{2010a}]{Vegetti10a}
{Vegetti} S.,  {Czoske} O.,   {Koopmans} L.~V.~E.,  2010a, \mn@doi [\mnras]
  {10.1111/j.1365-2966.2010.16952.x}, \href
  {http://adsabs.harvard.edu/abs/2010MNRAS.407..225V} {407, 225}

\bibitem[\protect\citeauthoryear{{Vegetti}, {Koopmans}, {Bolton}, {Treu}  \&
  {Gavazzi}}{{Vegetti} et~al.}{2010b}]{Vegetti10}
{Vegetti} S.,  {Koopmans} L.~V.~E.,  {Bolton} A.,  {Treu} T.,   {Gavazzi} R.,
  2010b, \mn@doi [\mnras] {10.1111/j.1365-2966.2010.16865.x}, \href
  {http://adsabs.harvard.edu/abs/2010MNRAS.408.1969V} {408, 1969}

\bibitem[\protect\citeauthoryear{{Vegetti}, {Lagattuta}, {McKean}, {Auger},
  {Fassnacht}  \& {Koopmans}}{{Vegetti} et~al.}{2012}]{Vegetti12}
{Vegetti} S.,  {Lagattuta} D.~J.,  {McKean} J.~P.,  {Auger} M.~W.,  {Fassnacht}
  C.~D.,   {Koopmans} L.~V.~E.,  2012, \mn@doi [\nat] {10.1038/nature10669},
  \href {http://adsabs.harvard.edu/abs/2012Natur.481..341V} {481, 341}

\bibitem[\protect\citeauthoryear{{Vegetti}, {Koopmans}, {Auger}, {Treu}  \&
  {Bolton}}{{Vegetti} et~al.}{2014}]{Vegetti14}
{Vegetti} S.,  {Koopmans} L.~V.~E.,  {Auger} M.~W.,  {Treu} T.,   {Bolton}
  A.~S.,  2014, \mn@doi [\mnras] {10.1093/mnras/stu943}, \href
  {http://adsabs.harvard.edu/abs/2014MNRAS.442.2017V} {442, 2017}

\bibitem[\protect\citeauthoryear{{Viel}, {Lesgourgues}, {Haehnelt}, {Matarrese}
   \& {Riotto}}{{Viel} et~al.}{2005}]{Viel05}
{Viel} M.,  {Lesgourgues} J.,  {Haehnelt} M.~G.,  {Matarrese} S.,   {Riotto}
  A.,  2005, \mn@doi [\prd] {10.1103/PhysRevD.71.063534}, \href
  {http://adsabs.harvard.edu/abs/2005PhRvD..71f3534V} {71, 063534}

\bibitem[\protect\citeauthoryear{{Vogelsberger}, {Zavala}, {Cyr-Racine},
  {Pfrommer}, {Bringmann}  \& {Sigurdson}}{{Vogelsberger} et~al.}{2016}]{Vog16}
{Vogelsberger} M.,  {Zavala} J.,  {Cyr-Racine} F.-Y.,  {Pfrommer} C.,
  {Bringmann} T.,   {Sigurdson} K.,  2016, \mn@doi [\mnras]
  {10.1093/mnras/stw1076}, \href
  {http://adsabs.harvard.edu/abs/2016MNRAS.460.1399V} {460, 1399}

\bibitem[\protect\citeauthoryear{{Walker} \& {Pe{\~n}arrubia}}{{Walker} \&
  {Pe{\~n}arrubia}}{2011}]{Walker2011}
{Walker} M.~G.,  {Pe{\~n}arrubia} J.,  2011, \mn@doi [\apj]
  {10.1088/0004-637X/742/1/20}, \href
  {http://adsabs.harvard.edu/abs/2011ApJ...742...20W} {742, 20}

\bibitem[\protect\citeauthoryear{{Watson}, {Li}  \& {Polley}}{{Watson}
  et~al.}{2012}]{Watson12}
{Watson} C.~R.,  {Li} Z.,   {Polley} N.~K.,  2012, \mn@doi [\jcap]
  {10.1088/1475-7516/2012/03/018}, \href
  {http://adsabs.harvard.edu/abs/2012JCAP...03..018W} {3, 018}

\bibitem[\protect\citeauthoryear{{Xu} et~al.,}{{Xu} et~al.}{2009}]{Xu09}
{Xu} D.~D.,  et~al., 2009, \mn@doi [\mnras] {10.1111/j.1365-2966.2009.15230.x},
  \href {http://adsabs.harvard.edu/abs/2009MNRAS.398.1235X} {398, 1235}

\bibitem[\protect\citeauthoryear{{Xu}, {Sluse}, {Gao}, {Wang}, {Frenk}, {Mao},
  {Schneider}  \& {Springel}}{{Xu} et~al.}{2015}]{Xu15}
{Xu} D.,  {Sluse} D.,  {Gao} L.,  {Wang} J.,  {Frenk} C.,  {Mao} S.,
  {Schneider} P.,   {Springel} V.,  2015, \mn@doi [\mnras]
  {10.1093/mnras/stu2673}, \href
  {http://adsabs.harvard.edu/abs/2015MNRAS.447.3189X} {447, 3189}

\bibitem[\protect\citeauthoryear{{de Blok}}{{de Blok}}{2010}]{deBlok2010}
{de Blok} W.~J.~G.,  2010, \mn@doi [Advances in Astronomy]
  {10.1155/2010/789293}, \href
  {http://adsabs.harvard.edu/abs/2010AdAst2010E...5D} {2010}

\makeatother
\end{thebibliography}

\bsp	
\label{lastpage}
\end{document}